\documentclass[american,brazil,english]{revtex4}
\usepackage[T1]{fontenc}
\usepackage[latin9]{inputenc}
\usepackage[a4paper]{geometry}
\usepackage{amsmath}
\usepackage{amssymb}
\usepackage{graphicx}
\usepackage{esint}

\makeatletter
\@ifundefined{textcolor}{}
{%
 \definecolor{BLACK}{gray}{0}
 \definecolor{WHITE}{gray}{1}
 \definecolor{RED}{rgb}{1,0,0}
 \definecolor{GREEN}{rgb}{0,1,0}
 \definecolor{BLUE}{rgb}{0,0,1}
 \definecolor{CYAN}{cmyk}{1,0,0,0}
 \definecolor{MAGENTA}{cmyk}{0,1,0,0}
 \definecolor{YELLOW}{cmyk}{0,0,1,0}
}

\usepackage{slashed}

\makeatother

\usepackage{babel}
\begin{document}
\selectlanguage{american}%

\title{Dynamical (super)symmetry vacuum properties of the supersymmetric
Chern-Simons-matter model}

\author{E. A. Gallegos }

\email{gallegos@fma.if.usp.br}

\selectlanguage{american}%

\author{A. J. da Silva}

\email{ajsilva@fma.if.usp.br}

\selectlanguage{american}%

\address{\textit{Instituto de Física, Universidade de São Paulo,}\\
 \textit{Caixa Postal 66318, 05315-970, São Paulo, SP, Brazil.} \\
}

\maketitle
By computing the two-loop effective potential of the $D=3$ $\mathcal{N}=1$
supersymmetric Chern-Simons model minimally coupled to a massless
self-interacting matter superfield, it is shown that supersymmetry
is preserved, while the internal $U(1)$ and the scale symmetries
are broken at two-loop order, dynamically generating masses both for
the gauge superfield and for the real component of the matter superfield.

\section{INTRODUCTION}

One of the main reasons for incorporating supersymmetry (susy) in
realistic quantum field theories (the standard model of particle physics)
is that this solves the gauge hierarchy problem, stabilizing the Higgs
mass against quadratic radiative corrections. However, since supersymmetry
has not been observed in Nature so far, it must be realized only in
its broken form. In this context dynamical supersymmetry breaking
(DSB), a beautiful phenomenon that occurs when the supersymmetry of
the vacuum at tree-level is broken by dynamical (perturbative or non-perturbative)
effects, has a privileged place in today's physics. Indeed, DSB not
only explains the stability of the Higgs boson, but also the origin
of the small mass ratios in the theory \cite{Witten-1981}. In four
dimensions (4D) DSB by perturbative effects (also known as Coleman-Weinberg's
mechanism) is forbidden by nonrenormalization theorems \cite{Grisaru-etal-1979}.
These theorems state that if supersymmetry is unbroken at tree level,
then it remains so to all orders in perturbation theory. DSB therefore
can only occur in 4D by nonperturbative effects (instantons, for example).

The nonexistence of such theorems in three dimensions (3D), in contrast,
opens the door for investigating this phenomenon owing to radiative
corrections in 3D supersymmetric field theories. In this paper, in
particular, we study the dynamical (super)symmetry properties of the
vacuum of the three dimensional $\mathcal{N}=1$ susy Chern-Simons
model minimally coupled to a massless self interacting matter field
(SCSM$_{3}$). 

Our interest in this kind of models is motivated in part by their
involvement in the construction of more complicated theories such
as the Bagger-Lambert-Gustavsson (BLG) theory \cite{BLG} and the
Aharony-Bergman-Jafferis-Maldacena (ABJM) theory \cite{ABJM} in connection
with the AdS$_{4}$/CFT$_{3}$ correspondence. In fact, in \cite{Raamsdonk}
and \cite{Keto-Kobayashi} it was shown that the BLG/ABJM theory in
terms of 3D $\mathcal{N}=1$ superfields \cite{Mauri-etal} involves
two non-Abelian supersymmetric Chern-Simons fields with opposite signs
and matter fields in the fundamental representation of the groups,
coupled to the two Chern-Simons fields (bifundamental matter). Moreover,
3D gauge Chern-Simons theories are important in their own right, as
they exhibit some remarkable features such as their topological nature
\cite{Deser-etal-topology} (quantization of the Chern-Simons coupling
constant) and their link with three dimensions through the $\epsilon_{\mu\nu\rho}$-tensor.
As far as physical applications are concerned, they play a significant
role in condensed-matter phenomena, e. g., in quantum Hall effect
\cite{Hall-Effect} and high-T$_{c}$ superconductivity \cite{Superconductivity}.

In this paper the behavior of the vacuum in SCSM$_{3}$ under radiative
corrections has been investigated by analyzing the minimum (or minima)
of the effective potential computed up to two loops in the superfield
perturbative formalism. The one-loop correction to the effective potential
was calculated by the tadpole method \cite{weinberg-1973}, while
the two-loop correction was calculated by the vacuum bubble method
\cite{jackiw}. Since in both methods the scalar superfields must
be shifted by their $\theta$ dependent vacuum expectation values,
we have to face the difficulty of dealing with an explicit breakdown
of supersymmetry in the intermediate stages of the calculation. Fortunately,
the projection operator method developed in \cite{boldo-helayel}
and recently extended in \cite{gallegos-adilson} allows us to derive
the supergraph Feynman rules, in particular, the superpropagators
for the broken susy theory. With this method, each superpropagator
of the shifted theory is expressed in terms of a basis of operators
in the respective sector.

The paper is structured as follows. In Sec. \ref{sec:Sec1} the three-dimensional
$\mathcal{N}=1$ supersymmetric Chern-Simons model coupled to matter
is introduced in the superfield formalism and its corresponding shifted
theory is constructed. The superpropagators of the shifted theory
are derived via the projection operator method. In Sec. \ref{sec: eff-pot}
the evaluation of the effective potential (in the Landau gauge $\alpha\rightarrow0$
and $\sigma_{2}$-linear approximation) is carried out by means of
the tadpole method and the vacuum bubble method. As argued in the
body of the paper these approximations are sufficient for our purposes.
The Appendices contain some details of the calculations.

\section{SETUP AND THE SCSM$_{3}$ MODEL \label{sec:Sec1}}

In the $D=3$ $\mathcal{N}=1$ superfield formalism, the building
blocks of supersymmetric Abelian gauge theories are (1) a complex
scalar (matter) superfield $\Phi\left(x,\,\theta\right)$ and (2)
a spinor gauge potential $A_{\alpha}\left(x,\,\theta\right)$. Adopting
the notation of \cite{Gates-etal}, the component-field contents of
these superfields are given by
\begin{equation}
\Phi(x,\,\theta)=\varphi\left(x\right)+\theta^{\alpha}\psi_{\alpha}\left(x\right)-\theta^{2}F\left(x\right)
\end{equation}
and
\begin{equation}
A_{\alpha}\left(x,\,\theta\right)=\chi_{\alpha}\left(x\right)-\theta_{\alpha}B(x)+i\theta^{\beta}V_{\alpha\beta}\left(x\right)-\theta^{2}\left(2\lambda_{\alpha}+i\partial_{\alpha\beta}\chi^{\beta}\right).
\end{equation}

Using these superfields along with the supersymmetric gauge covariant
derivative $\nabla_{\alpha}\doteq D_{\alpha}-ieA_{\alpha}$, with
$D_{\alpha}\doteq\partial_{\alpha}+i\theta^{\beta}\partial_{\alpha\beta}$,
the three-dimensional $\mathcal{N}=1$ supersymmetric Chern-Simons
model coupled to matter (SCSM$_{3}$) is described by the action
\begin{equation}
S=\int d^{5}z\left\{ A^{\alpha}W_{\alpha}-\frac{1}{2}\overline{\nabla}^{\alpha}\overline{\Phi}\nabla_{\alpha}\Phi-g\left(\overline{\Phi}\Phi\right)^{2}\right\} ,\label{eq:2.1}
\end{equation}
where $W_{\alpha}\doteq\frac{1}{2}D^{\beta}D_{\alpha}A_{\beta}$ is
the superfield strength that satisfies the Bianchi identity $D^{\alpha}W_{\alpha}=0$.

The action (\ref{eq:2.1}) is invariant under the following infinitesimal
gauge transformations
\begin{equation}
\Phi'=\left(1+ieK\right)\Phi,\qquad A'_{\alpha}=A_{\alpha}+D_{\alpha}K,\label{eq:2.2}
\end{equation}
with $K\left(x,\,\theta\right)$ denoting an arbitrary real scalar
superfield,
\begin{equation}
K\left(x,\,\theta\right)=\omega\left(x\right)+\theta^{\alpha}\sigma_{\alpha}\left(x\right)-\theta^{2}\tau\left(x\right).\label{eq:2.2-1}
\end{equation}
 Notice that under these transformations the superfield strength $W_{\alpha}$
is invariant $\left(W'_{\alpha}=W_{\alpha}\right)$, whereas the derivative
$\nabla_{\alpha}$ transforms like a covariant object, namely, $\nabla'_{\alpha}\Phi'=\left(1+ieK\right)\nabla_{\alpha}\Phi$.

Since our purpose is to calculate the two-loop effective potential
by means of the tadpole method \cite{weinberg-1973} at one-loop order
and the vacuum bubble method \cite{jackiw} at two-loop order, we
must appropriately choose the gauge fixing term in order to quantize
the theory. The simplest choice compatible with both methods is the
Lorentz-like gauge fixing term,
\begin{eqnarray}
S_{FG} & = & \int d^{5}z\,\left(-\frac{1}{4\alpha}\right)D^{\alpha}A_{\alpha}D^{\beta}A_{\beta},\label{eq:2.3}
\end{eqnarray}
where $\alpha$ is a dimensionless parameter. The advantage of fixing
the gauge in this way is that the Faddeev-Popov ghosts remain free
and can be ignored.

Writing the complex matter superfield $\Phi$ in terms of two real
superfields $\Sigma$ and $\Pi$,
\begin{equation}
\Phi=\frac{1}{\sqrt{2}}\left(\Sigma+i\Pi\right),\label{eq:2.4}
\end{equation}
and adding (\ref{eq:2.3}) to (\ref{eq:2.1}), the classical action
reads
\begin{eqnarray}
S & = & \int d^{5}z\left\{ \frac{1}{2}A_{\alpha}\left[-D^{\beta}D^{\alpha}-\frac{1}{2\alpha}D^{\alpha}D^{\beta}\right]A_{\beta}+\frac{1}{2}\Sigma D^{2}\Sigma+\frac{1}{2}\Pi D^{2}\Pi-\frac{e}{2}D^{\alpha}\Sigma A_{\alpha}\Pi\right.\nonumber \\
 &  & \left.+\frac{e}{2}D^{\alpha}\Pi A_{\alpha}\Sigma-\frac{e^{2}}{2}\left(\Sigma^{2}+\Pi^{2}\right)A^{2}-\frac{g}{4}\left(\Sigma^{2}+\Pi^{2}\right)^{2}\right\} .\label{eq:2.5}
\end{eqnarray}

A simple dimensional analysis ($c=\hbar=1$) shows that all theory's
parameters, i.e. $\alpha$, \textbf{$e$} and $g$ are dimensionless.
Hence this model is a kind of 3D susy version of the conformally invariant
Coleman-Weinberg model \cite{Coleman-Weinberg} (for this reason it
is sometimes called, in the literature, the 3D susy Coleman-Weinberg
model). Furthermore, it should be noted that the quadratic term in
the gauge superfield $A_{\alpha}$ is not the Maxwell term, but instead
the well-known Chern-Simons term
\begin{equation}
-\int d^{5}z\frac{1}{2}A_{\alpha}D^{\beta}D^{\alpha}A_{\beta}=-\int d^{3}x\frac{1}{2}\epsilon^{\mu\nu\rho}v_{\mu}\partial_{\nu}v_{\rho}+\cdots,
\end{equation}
where the ellipsis represents other terms, $\epsilon^{\mu\nu\rho}$
$\left(\epsilon^{012}\doteq1\right)$ is the completely antisymmetric
tensor in the Minkowski space, and $v_{\mu}$ is the three vector
given by $v_{\mu}\doteq\left(\gamma_{\mu}\right)^{\alpha\beta}V_{\alpha\beta}$.

In order to compute the effective potential by using the tadpole \cite{weinberg-1973}
and the vacuum bubble \cite{jackiw} methods, we must shift in (\ref{eq:2.5})
both scalar superfields ($\Sigma$, $\Pi$):
\begin{equation}
\Sigma\rightarrow\Sigma+\sigma\left(\theta\right),\qquad\Pi\rightarrow\Pi+\pi\left(\theta\right),\label{eq:2.6}
\end{equation}
where $\sigma\left(\theta\right)\doteq\sigma_{1}-\theta^{2}\sigma_{2}$
and $\pi\left(\theta\right)\doteq\pi_{1}-\theta^{2}\pi_{2}$, with
$\sigma_{i}$ and $\pi_{i}$ being $x$-constant classical fields
($\sigma_{1}$ and $\pi_{1}$are dynamical component fields and $\sigma_{2}$and
$\pi_{2}$ are auxiliary fields, whose non-null values imply in breakdown
of susy in the intermediate steps of the calculations). However, we
can make use of the rotational $SO(2)$ symmetry, $\sigma'_{i}+i\pi'_{i}=\exp\left(i\, e\,\omega\right)\left(\sigma_{i}+i\pi_{i}\right)$,
that the effective potential inherits from the classical action, to
simplify the calculations. By taking advantage of this symmetry we
will only shift the real scalar superfield $\Sigma$. At the end of
calculations, for the analysis of the results, the rotational symmetry
$SO(2)$ will be restored by performing the following substitutions:
\begin{equation}
\sigma_{i}^{2}\rightarrow\sigma_{i}^{2}+\pi_{i}^{2},\qquad\sigma_{1}\sigma_{2}\rightarrow\sigma_{1}\sigma_{2}+\pi_{1}\pi_{2}
\end{equation}
After performing the $\Sigma$ shift in (\ref{eq:2.5}), the shifted
action $S'$ may be written as
\begin{align}
S' & =\int d^{5}zd^{5}z'\biggl[\frac{1}{2}A_{\alpha}\left(z\right)\mathcal{O}^{\alpha\beta}\left(z,\, z'\right)A_{\beta}\left(z'\right)+\frac{1}{2}\Sigma\left(z\right)\mathcal{O}^{(\Sigma)}\left(z,\, z'\right)\Sigma\left(z'\right)+\frac{1}{2}\Pi\left(z\right)\mathcal{O}^{(\Pi)}\left(z,\, z'\right)\Pi\left(z'\right)\nonumber \\
 & +A^{\alpha}\left(z\right)\mathcal{O}_{\alpha}\left(z,\, z'\right)\Pi\left(z'\right)\biggr]+\int d^{5}z\biggl[-\frac{e}{2}\left(D^{\alpha}\Sigma\Pi-D^{\alpha}\Pi\Sigma\right)A_{\alpha}-e^{2}\sigma\left(\theta\right)\Sigma A^{2}\nonumber \\
 & -g\sigma\left(\theta\right)\left(\Sigma^{3}+\Sigma\Pi^{2}\right)-\frac{e^{2}}{2}\left(\Sigma^{2}+\Pi^{2}\right)A^{2}-\frac{g}{4}\left(\Sigma^{2}+\Pi^{2}\right)^{2}+\left(D^{2}\sigma-g\sigma^{3}\right)\Sigma+\frac{1}{2}\sigma D^{2}\sigma-\frac{g}{4}\sigma^{4}\biggr],\label{eq:2.7}
\end{align}
where we have introduced the supermatrices
\begin{eqnarray}
\mathcal{O}_{\alpha\beta}\left(z,\, z'\right) & = & \left[-D_{\beta}D_{\alpha}-\frac{1}{2\alpha}D_{\alpha}D_{\beta}+\frac{e^{2}}{2}\sigma^{2}\left(\theta\right)C_{\alpha\beta}\right]\delta^{5}\left(z-z'\right)\nonumber \\
\mathcal{O}^{(\Sigma)}\left(z,\, z'\right) & = & \left[D^{2}-3g\sigma^{2}\left(\theta\right)\right]\delta^{5}\left(z-z'\right)\nonumber \\
\mathcal{O}^{(\Pi)}\left(z,\, z'\right) & = & \left[D^{2}-g\sigma^{2}\left(\theta\right)\right]\delta^{5}\left(z-z'\right)\nonumber \\
\mathcal{O}_{\alpha}\left(z,\, z'\right) & = & \left[\frac{e}{2}\left(\sigma\left(\theta\right)D_{\alpha}-D_{\alpha}\sigma\left(\theta\right)\right)\right]\delta^{5}\left(z-z'\right).\label{eq:2.7-1}
\end{eqnarray}
From these equations, as we shall see below, the superpropagators
of the shifted theory are calculated. Linear terms in $\Sigma$ and
$x$-constant terms are retained in the action because they define
the $\Sigma$-tadpole and the vacuum bubble at tree level, respectively.
Moreover, from now on we will assume that the vacuum expectation values
of the new scalar superfields are zero: $\left\langle \Sigma\right\rangle =\left\langle \Pi\right\rangle =0$.

As it can be seen from the above action (\ref{eq:2.7}), the effect
of the shift is to induce ``masses'' for the scalar superfields
$\left(\Sigma,\,\Pi\right)$. Due to the non-null value of the auxiliary
field $\sigma_{2}$ the mass of the scalar and the fermionic components
of each superfield are different and susy is broken (this fact can
be explicitly seen by calculating the component field propagators
of the superfields). Another effect is the induction of a mixing between
$A_{\alpha}$ and $\Pi$. It is worth mentioning that this mixture
is unavoidable (when the classical auxiliary fields $\sigma_{2}$
and/or $\pi_{2}$ are non-null) even if one employs an extension of
the $R_{\xi}$ gauge. So, in the intermediate stages of the calculation,
the non-null $\sigma_{2}$ auxiliary field implies in the breakdown
of susy, giving different masses for the bosonic and fermionic components
of the superfields (as we will see at the end of the calculation,
the minimum of the effective potential, in fact occurs for $\sigma_{2}=0=\pi_{2}$,
implying in the conservation of susy).

Before starting with the calculation of the effective potential up
to two-loop order, it is necessary to establish the supergraph Feynman
rules for the shifted theory (\ref{eq:2.7}), in particular to derive
its shifted superpropagators. As is usual in quantum field theory,
they are derived by explicitly integrating the free generating functional
$\mathcal{Z}_{0}\left[J,\, G\,\eta\right]$ of the shifted theory,
\begin{equation}
\mathcal{Z}_{0}\left[J,\, G,\,\eta\right]=\mathcal{N}\,\int\mathcal{D}\Sigma\,\mathcal{D}\Pi\,\mathcal{D}A_{\alpha}\exp i\left\{ S_{bil}\left[\Sigma,\,\Pi,\, A_{\alpha}\right]+J\cdot\Sigma+G\cdot\Pi+\eta^{\alpha}\cdot A_{\alpha}\right\} ,
\end{equation}
where $S_{bil}$ stands for the bilinear part of the shifted action
(\ref{eq:2.7}) and $\left\{ J,\, G,\,\eta^{\alpha}\right\} $ are
external sources for $\Sigma$, $\Pi$ and $A_{\alpha}$, respectively.
In addition, the dot mark in $X\cdot Y$ means $X\cdot Y\doteq\int d^{5}zX\left(z\right)Y\left(z\right)$.

In this way after taking the appropriate functional derivatives of
the integrated free functional $\mathcal{Z}_{0}\left[J,\, G\,\eta\right]$,
the shifted superpropagators are given by
\begin{eqnarray}
\left\langle T\, A_{\alpha}\left(z\right)A_{\beta}\left(z^{\prime}\right)\right\rangle  & = & i\,\Theta^{-1}{}_{\alpha\beta}\left(z,\, z^{\prime}\right)\nonumber \\
\left\langle T\,\Pi\left(z\right)\Pi\left(z^{\prime}\right)\right\rangle  & = & i\,\mathcal{O}^{(\Pi)-1}\left(z,\, z^{\prime}\right)+i\int\int_{z_{1},\, z_{2}}\mathcal{O}^{(\Pi)-1}\left(z,\, z_{1}\right)H\left(z_{1},\, z_{2}\right)\mathcal{O}^{(\Pi)-1}\left(z_{2},\, z^{\prime}\right)\nonumber \\
\left\langle T\,\Pi\left(z\right)A_{\alpha}\left(z^{\prime}\right)\right\rangle  & = & -i\int\int_{z_{1},\, z_{2}}\ \mathcal{O}^{(\Pi)-1}\left(z,\, z_{1}\right)\mathcal{O}^{\beta}\left(z_{2},\, z_{1}\right)\Theta^{-1}{}_{\beta\alpha}\left(z_{2},\, z'\right)\nonumber \\
\left\langle T\,\Sigma\left(z\right)\Sigma\left(z^{\prime}\right)\right\rangle  & = & i\,\mathcal{O}^{(\Sigma)-1}\left(z,\, z^{\prime}\right),\label{eq:2.8}
\end{eqnarray}
with
\begin{eqnarray}
\Theta_{\alpha\beta}\left(z,\, z^{\prime}\right) & = & \mathcal{O}_{\alpha\beta}\left(z,\, z^{\prime}\right)+Q_{\alpha\beta}\left(z,\, z^{\prime}\right)\nonumber \\
Q_{\alpha\beta}\left(z,\, z^{\prime}\right) & = & \int\int_{z_{1},z_{2}}\mathcal{O}_{\alpha}\left(z,\, z_{1}\right)\mathcal{O}^{(\Pi)-1}\left(z_{1},\, z_{2}\right)\mathcal{O}_{\beta}\left(z^{\prime},\, z_{2}\right)\nonumber \\
H\left(z,\, z^{\prime}\right) & = & \int\int_{z_{1},z_{2}}\mathcal{O}^{\alpha}\left(z_{1},\, z\right)\Theta^{-1}{}_{\alpha}^{\hspace{6pt}\beta}\left(z_{1},\, z_{2}\right)\mathcal{O}_{\beta}\left(z_{2},\, z^{\prime}\right).\label{eq:2.8-1}
\end{eqnarray}
From these expressions one sees that the gauge-scalar mixture in (\ref{eq:2.7})
has two effects. Firstly, this gives rise to a mixing propagator between
$\Pi$ and $A_{\alpha}$, and secondly, it changes the pure superpropagators
for $A_{\alpha}$ and $\Pi$ which the theory would have without the
presence of the mixture.

By carrying out all the algebraic operations involved in (\ref{eq:2.8}-\ref{eq:2.8-1})
through the projection operators method developed in \cite{boldo-helayel}
and recently enlarged (in the gauge sector) in \cite{gallegos-adilson},
the superpropagators of the shifted theory can be written as \begin{subequations}
\begin{align}
\left\langle T\, A_{\alpha}\left(k,\theta\right)A_{\beta}\left(-k,\theta^{\prime}\right)\right\rangle  & =i\left\{ \mbox{\ensuremath{\ensuremath{\sum\limits _{i=0}^{5}\left(r_{i}R_{i,\alpha\beta}+s_{i}S_{i,\alpha\beta}\right)}+}\ensuremath{m}}M_{\alpha\beta}+nN_{\alpha\beta}\right\} \delta^{2}\left(\theta-\theta^{\prime}\right)\\
\left\langle T\,\Pi\left(k,\,\theta\right)\Pi\left(-k,\,\theta^{\prime}\right)\right\rangle  & =i\,\left(\sum\limits _{i=0}^{5}a_{i}\, P_{i}\right)\delta^{2}\left(\theta-\theta^{\prime}\right)\\
\left\langle T\,\Pi\left(k,\,\theta\right)A_{\alpha}\left(-k,\,\theta^{\prime}\right)\right\rangle  & =i\left(\sum_{i=1}^{8}b_{i}\, T_{\alpha}^{i}\right)\delta^{2}\left(\theta-\theta^{\prime}\right)\\
\left\langle T\,\Sigma\left(k,\,\theta\right)\Sigma\left(-k,\,\theta^{\prime}\right)\right\rangle  & =i\,\left(\sum\limits _{i=0}^{5}c_{i}\, P_{i}\right)\delta^{2}\left(\theta-\theta^{\prime}\right),
\end{align}
\label{propagators}\end{subequations}where the set
\begin{equation}
P_{0}\doteq1,\,\, P_{1}\doteq D^{2},\,\, P_{2}\doteq\theta^{2},\,\, P_{3}\doteq\theta^{\alpha}D_{\alpha},\,\, P_{4}\doteq\theta^{2}D^{2},\,\, P_{5}\doteq k_{\alpha\beta}\theta^{\alpha}D^{\beta}\label{eq:2.9a}
\end{equation}
forms an operator basis in the scalar sector, the set
\begin{equation}
R_{i}^{\alpha\beta}\doteq k^{\alpha\beta}P_{i},\, S_{i}^{\alpha\beta}\doteq C^{\alpha\beta}P_{i},\, M^{\alpha\beta}\doteq\theta^{\alpha}D^{\beta}+\theta^{\beta}D^{\alpha},\, N^{\alpha\beta}\doteq k^{\alpha\gamma}\theta^{\beta}D_{\gamma}+k^{\beta\gamma}\theta^{\alpha}D_{\gamma},\label{eq:2.9b}
\end{equation}
an operator basis in the gauge sector and the set
\begin{equation}
\begin{array}{cccccccccc}
T_{\alpha}^{1}\doteq\theta_{\alpha}, &  &  & T_{\alpha}^{2}\doteq k_{\alpha\beta}\theta^{\beta}, &  &  & T_{\alpha}^{3}\doteq\theta_{\alpha}D^{2}, &  &  & T_{\alpha}^{4}\doteq k_{\alpha\beta}\theta^{\beta}D^{2},\\
\\
T_{\alpha}^{5}\doteq D_{\alpha}, &  &  & T_{\alpha}^{6}\doteq k_{\alpha\beta}D^{\beta}, &  &  & T_{\alpha}^{7}\doteq\theta^{2}D_{\alpha}, &  &  & T_{\alpha}^{8}\doteq k_{\alpha\beta}\theta^{2}D^{\beta},
\end{array}\label{eq:2.9c}
\end{equation}
an operator basis in the mixing sector. For more details about these
bases the reader is referred to \cite{gallegos-adilson}.

The coefficients $r_{i},\cdots,c_{i}$ in the $\left(\alpha,\,\sigma_{2}\right)$-linear
approximation are collected in Appendix \ref{sec:App-Coeff}. These
approximations are sufficient to study the vacuum properties of the
SCSM$_{3}$ model. Indeed, the $\sigma_{2}$-linear approximation
as discussed in \cite{LA-Gaume} and reproduced in our paper \cite{gallegos}
is enough to study the possibility of susy breaking by radiative corrections,
while the $\alpha$-linear approximation (taking the Landau gauge
$\alpha\rightarrow0$ in the final stage) is merely a technical one
since the coefficients for a generic gauge parameter are very intricate.
Nevertheless, even though the effective potential of gauge theories
is a gauge-dependent quantity \cite{dolan-jackiw} (explicitly dependent
of the gauge parameter $\alpha$), its vacuum properties are gauge
independent, as assured by the Nielsen identities \cite{nielsen}.

\section{THE EFFECTIVE POTENTIAL UP TO TWO-LOOPS\label{sec: eff-pot}}

In what follows we are going to compute the two-loop contribution
to the effective potential of the SCSM$_{3}$ model. The classical
potential is defined (in the vacuum bubble method) by the $x$-constant
terms which appear in (\ref{eq:2.7}), that is,
\begin{eqnarray}
U_{cl}\left(\sigma_{1},\,\sigma_{2}\right) & = & -\int d^{2}\theta\left\{ \frac{1}{2}\sigma\left(\theta\right)D^{2}\sigma\left(\theta\right)-\frac{g}{4}\sigma^{4}\left(\theta\right)\right\} \nonumber \\
 & = & -\frac{1}{2}\sigma_{2}^{2}+g\sigma_{1}^{3}\sigma_{2},\label{eq:3.1}
\end{eqnarray}
where an overall spacetime factor $\left(\int d^{3}x\right)$ was
dropped. Solving the Euler-Lagrange equation for $\sigma_{2}$ we
get $\sigma_{2}(\sigma_{1})=g\sigma_{1}^{3}$ and $U_{cl}(\sigma_{1})=g^{2}\sigma_{1}^{6}/2.$
For future use we write the two expressions for $U_{cl}$ after restoring
the rotational symmetry in the scalar superfields. The results are
\begin{equation}
U_{cl}\left(\sigma_{i},\,\pi_{i}\right)=-\frac{1}{2}(\sigma_{2}^{2}+\pi_{2}^{2})+g(\sigma_{1}^{2}+\pi_{1}^{2})(\sigma_{1}\sigma_{2}+\pi_{1}\pi_{2})\label{eq:3.1-1}
\end{equation}
and
\begin{equation}
U_{cl}\left(\sigma_{1},\,\pi_{1}\right)=\frac{g^{2}}{2}(\sigma_{1}^{2}+\pi_{1}^{2})^{3}.\label{eq:3.1-2}
\end{equation}
The above results can also be achieved by using the tadpole method.
In this case, the tree-level $\Sigma$ supertadpole is read directly
from (\ref{eq:2.7}),
\begin{eqnarray}
\Gamma_{cl}^{\left(\Sigma\right)} & = & \int d^{3}xd^{2}\theta\left(D^{2}\sigma-g\sigma^{3}\right)\Sigma\left(x,\,\theta\right)\nonumber \\
 & = & \int d^{3}x\left[-3g\sigma_{1}^{2}\sigma_{2}\Sigma_{1}\left(x\right)+\left(\sigma_{2}-g\sigma_{1}^{3}\right)\Sigma_{2}\left(x\right)\right],
\end{eqnarray}
where the second line results from integrating over $\theta$, using
the fact that $\Sigma\left(x,\,\theta\right)\doteq\Sigma_{1}\left(x\right)+\theta^{\alpha}\Psi_{\alpha}\left(x\right)-\Sigma_{2}\left(x\right)\theta^{2}$.
Identifying the tree-level $\Sigma_{1}\left(\Sigma_{2}\right)$ tadpoles
from this last expression, it is straightforward to set up the tadpole
equations
\begin{eqnarray}
\frac{\partial U_{cl}}{\partial\sigma_{1}} & = & 3g\sigma_{1}^{2}\sigma_{2},\\
\frac{\partial U_{cl}}{\partial\sigma_{2}} & = & -\left(\sigma_{2}-g\sigma_{1}^{3}\right),
\end{eqnarray}
which in turn consistently provide the same solution as before: $U_{cl}=-\frac{1}{2}\sigma_{2}^{2}+g\sigma_{1}^{3}\sigma_{2}$.

\selectlanguage{english}%

\begin{figure}
\begin{centering}
\includegraphics[scale=0.4]{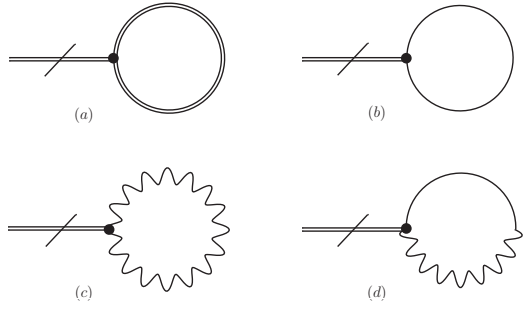}
\par\end{centering}

\caption{\selectlanguage{brazil}%
\label{fig:1-loop}One-loop $\Sigma$ supertadpoles of the shifted
SCSM$_{3}$ model. Double-solid lines represent scalar $\Sigma$ propagators,
the solid line represents the scalar $\Pi$ propagator, the wavy line
represents the gauge propagator and the solid-wavy line represents
the mixing $\left\langle \Pi A\right\rangle $ propagator.\selectlanguage{american}%
}
\end{figure}

\selectlanguage{american}%
In the one-loop level the $\Sigma$ supertadpoles that contribute
to the effective action are shown in Figure \ref{fig:1-loop}. Their
corresponding integrals are given by
\begin{eqnarray}
\Gamma_{1}^{(\Sigma)} & = & \int d\widetilde{p}\frac{d^{3}k}{\left(2\pi\right)^{3}}\int d^{2}\theta\Bigl[-3g\sigma\left(\theta\right)\left\langle \Sigma\left(k,\theta\right)\Sigma\left(-k,\theta\right)\right\rangle -g\sigma\left(\theta\right)\left\langle \Pi\left(k,\theta\right)\Pi\left(-k,\theta\right)\right\rangle \nonumber \\
 &  & -\frac{e^{2}}{2}\sigma\left(\theta\right)\left\langle A^{\alpha}\left(k,\theta\right)A_{\alpha}\left(-k,\theta\right)\right\rangle +e\left\langle D^{\alpha}\Pi\left(k,\theta\right)A_{\alpha}\left(-k,\theta\right)\right\rangle \nonumber \\
 &  & +\frac{e}{2}\left\langle D^{\alpha}A_{\alpha}\left(k,\theta\right)\Pi\left(-k,\theta\right)\right\rangle \Bigr]\widetilde{\Sigma}\left(p,\,\theta\right),
\end{eqnarray}
with $d\widetilde{p}\doteq\frac{d^{3}p}{\left(2\pi\right)^{3}}\left(2\pi\right)^{3}\delta^{2}\left(p\right)$.

Inserting the superpropagators (\ref{propagators}) into the expression
above and integrating over $\theta$, one obtains
\begin{eqnarray}
\Gamma_{1}^{(\Sigma)} & = & i\int d^{3}\widetilde{p}\frac{d^{3}k}{\left(2\pi\right)^{3}}\Bigl[\Bigl(e^{2}\sigma_{1}s_{1}\left(k\right)+eb_{5}\left(k\right)-2eb_{3}\left(k\right)-g\sigma_{1}a_{1}\left(k\right)-3g\sigma_{1}c_{1}\left(k\right)\Bigr)\widetilde{\Sigma}_{2}(p)+\nonumber \\
 &  & -\Bigl(-e^{2}\sigma_{2}s_{1}\left(k\right)+e^{2}\sigma_{1}s_{4}\left(k\right)+eb_{7}\left(k\right)+g\sigma_{2}a_{1}\left(k\right)-g\sigma_{1}a_{4}\left(k\right)\nonumber \\
 &  & +3g\sigma_{2}c_{1}\left(k\right)-3g\sigma_{1}c_{4}\left(k\right)\Bigr)\widetilde{\Sigma}_{1}(p)\Bigr].\label{eq:3.2}
\end{eqnarray}
It is important to note that it was not necessary to consider the
explicit form of the propagator coefficients in order to perform the
Grassmann integration (i.e. the D-algebra). This is always possible
since the propagator coefficients are merely functions on $k^{2}$
and the parameters of the shifted theory, while the D-algebra entails
$(\theta_{\alpha},\, D_{\alpha},\, k_{\alpha\beta})$ manipulations
which are explicit in the definitions of the bases (\ref{eq:2.9a}-\ref{eq:2.9c}).

To proceed, as was made in the tree-level case, we set up the tadpole
equations by reading directly the $\Sigma_{1}\left(\Sigma_{2}\right)$
tadpoles from (\ref{eq:3.2}). This leads to
\begin{eqnarray}
\frac{\partial U_{1}}{\partial\sigma_{1}} & = & i\int\frac{d^{3}k}{\left(2\pi\right)^{3}}\left[-e^{2}\sigma_{2}s_{1}\left(k\right)+e^{2}\sigma_{1}s_{4}\left(k\right)+eb_{7}\left(k\right)+g\sigma_{2}a_{1}\left(k\right)-g\sigma_{1}a_{4}\left(k\right)\right.\nonumber \\
 &  & \left.+3g\sigma_{2}c_{1}\left(k\right)-3g\sigma_{1}c_{4}\left(k\right)\right]\\
\frac{\partial U_{1}}{\partial\sigma_{2}} & = & -i\int\frac{d^{3}k}{\left(2\pi\right)^{3}}\left[e^{2}\sigma_{1}s_{1}\left(k\right)+eb_{5}\left(k\right)-2eb_{3}\left(k\right)-g\sigma_{1}a_{1}\left(k\right)-3g\sigma_{1}c_{1}\left(k\right)\right],
\end{eqnarray}
where the coefficients $\left\{ a_{i},\, b_{i},\, c_{i},\, s_{i}\right\} $
are functions on $\sigma_{1}$ and $\sigma_{2}$ (see Appendix \ref{sec:App-Coeff}
).

Solving this system of differential equations, the one-loop contribution
(in the Landau gauge $\alpha=0$) is
\begin{eqnarray}
U_{1}\left(\sigma_{i}\right) & = & \frac{\sigma_{1}\sigma_{2}}{4}i^{2}\int\frac{d^{3}k_{E}}{\left(2\pi\right)^{3}}\frac{-3e^{2}g^{3}\sigma_{1}^{8}\left(e^{2}-3g\right)-g\sigma_{1}^{4}\left(e^{4}-10e^{2}g+48g^{2}\right)k_{E}^{2}+\left(e^{2}-16g\right)k_{E}^{4}}{\left(k_{E}^{2}+\mu_{1}^{2}\right)\left(k_{E}^{2}+\mu_{2}^{2}\right)\left(k_{E}^{2}+\mu_{3}^{2}\right)}\nonumber \\
 & = & \frac{1}{64\pi}\left(e^{4}-160g^{2}\right)\sigma_{1}^{3}\sigma_{2}+\mathcal{O}\left(\sigma_{1},\,\sigma_{2}^{2}\right).\label{eq:3.3}
\end{eqnarray}
Here $k_{E}$ represents the Euclidean momentum. As is seen from the
sum of (\ref{eq:3.1}) and (\ref{eq:3.3}), neither the supersymmetry
nor the internal $U\left(1\right)$ symmetry are broken up to this
order.

\begin{figure}[h]
\begin{centering}
\includegraphics[scale=0.65]{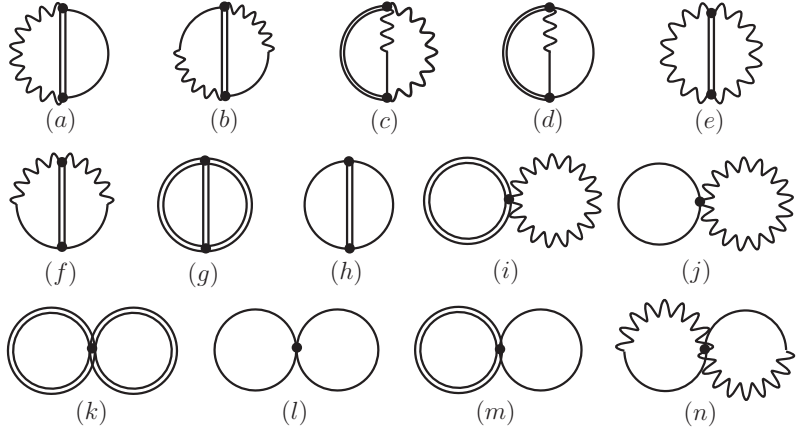}
\par\end{centering}

\caption{\selectlanguage{brazil}%
\label{fig:2-loop} Two-loop vacuum bubbles of the shifted SCSM$_{3}$
model.\selectlanguage{american}%
}
\end{figure}

Now let us go to the two-loop approximation. In this order, the vacuum
bubbles which contribute to the effective potential are displayed
in Figure \ref{fig:2-loop}. Their respective integrals after performing
the D-algebra, with the aid of the SusyMath package \cite{Ferrari},
are collected in Appendix \ref{sec:App-2-loops}.

Using dimensional regularization to integrate over the internal momenta
and specifically the formulas found in \cite{Tan-etal,dias-adilson},
we obtain for the two-loop contribution the following result 
\begin{eqnarray}
U_{2}\left(\sigma_{1},\sigma_{2}\right) & = & \frac{1}{512\pi^{2}}\left[\frac{2a_{1}}{\epsilon}+a_{2}-4a_{1}\ln\left(\frac{\sigma_{1}^{2}}{\mu}\right)\right]\sigma_{1}^{3}\sigma_{2}+B_{ct}\sigma_{1}^{3}\sigma_{2},\label{eq:3.4}
\end{eqnarray}
where $\epsilon=3-D$ and $\mu$ is an arbitrary mass scale introduced
in the dimensional regularization. The constant $B_{ct}$, chosen
as $B_{ct}=-\frac{2a_{1}}{512\pi^{2}}$$\frac{1}{\epsilon}+B_{fin}$
is a tree-level counterterm to the coupling constant $g$ that will
cancel the two-loop infinite and adjust the coupling constant to the
required renormalization condition. The constants $a_{1}$ and $a_{2}$
are given by
\begin{eqnarray}
a_{1} & \doteq & e^{6}+7e^{4}g-16e^{2}g^{2}-1024g^{3},
\end{eqnarray}
and
\begin{eqnarray}
a_{2} & \doteq & a_{1\,}2\left(1-\gamma+\ln(4\pi)\right)+32g^{3}\left[-47\ln2+243\ln3+20\left(5+\ln5\right)\right]+3936g^{3}\ln g\nonumber \\
 &  & +4e^{4}g\left(-5+\ln256\right)+\left(4\ln2-1\right)e^{6}-4\left(e^{2}-6g\right)\left(e^{2}+6g\right)^{2}\ln\left(e^{2}+6g\right)\nonumber \\
 &  & +16e^{2}g^{2}\left(\ln2-5\right)+\left[12e^{4}g-144e^{2}g^{2}-1728g^{3}+e^{6}\right]\ln\left(e^{2}+12g\right)\nonumber \\
 &  & -\left(e^{2}-8g\right)\left(e^{2}+8g\right)\left(e^{2}+16g\right)\ln\left(e^{2}+16g\right),
\end{eqnarray}
where $\gamma=0.5772\cdots$ is the Euler's constant. Defining the
constants\foreignlanguage{english}{
\begin{equation}
Y\left(e,g\right)\doteq\frac{a_{1}}{128\pi^{2}}\qquad\mbox{and}\qquad X\left(e,g\right)\doteq\frac{e^{4}-160g^{2}}{64\pi}+\frac{a_{2}}{512\pi^{2}}+B_{fin},
\end{equation}
}the effective potential up to two-loops (in the Landau gauge $\alpha\rightarrow0$
and $\sigma_{2}$-linear approximation in the loop corrections) is
given by the sum of (\ref{eq:3.1}), (\ref{eq:3.3}) and (\ref{eq:3.4})
\begin{eqnarray}
U\left(\sigma_{1},\,\sigma_{2}\right) & = & -\frac{1}{2}\sigma_{2}^{2}+\left[g+X\left(e,g\right)-Y\left(e,g\right)\ln\frac{\sigma_{1}^{2}}{\mu}\right]\sigma_{1}^{3}\sigma_{2}+\mathcal{O}\left(\sigma_{1},\,\sigma_{2}^{2}\right).\label{eq:3.5}
\end{eqnarray}

Eliminating the auxiliary field $\sigma_{2}$ by using its Euler-Lagrange
equation $\partial U/\partial\sigma_{2}=0$ we get\foreignlanguage{english}{
\begin{equation}
\sigma_{2}(\sigma_{1})=\left[g+X\left(e,g\right)-Y\left(e,g\right)\ln\frac{\sigma_{1}^{2}}{\mu}\right]\sigma_{1}^{3},\label{eq:3.6}
\end{equation}
which substituted in $U\left(\sigma_{1},\,\sigma_{2}\right)$ results
in
\begin{equation}
U(\sigma_{1})=\frac{1}{2}\sigma_{2}^{2}(\sigma_{1})=\frac{\sigma_{1}^{6}}{2}\left[g+X\left(e,g\right)-Y\left(e,g\right)\ln\frac{\sigma_{1}^{2}}{\mu}\right]^{2}.\label{eq:3.7}
\end{equation}
}

Besides the usual minimum at $\sigma_{1}=0$, this potential has a
possible new minimum at $\sigma_{1}=\eta\neq0$ satisfying $g+X\left(e,g\right)-Y\left(e,g\right)\ln\frac{\eta^{2}}{\mu}=0.$
By imposing the renormalization condition:
\begin{equation}
\left.\frac{\partial^{6}U}{\partial\sigma_{1}^{6}}\right|_{\sigma_{1}=\eta}=\frac{\partial^{6}U_{cl}}{\partial\sigma_{1}^{6}}=360g^{2},
\end{equation}
we obtain the relation
\begin{equation}
\sqrt{\frac{45}{812}}g=Y=\frac{1}{128\pi^{2}}(e^{6}+e^{4}g+\cdots)
\end{equation}
between the two coupling constants. Up to order $e^{6}\ll1,$ this
condition implies that $Y=\frac{e^{6}}{128\pi^{2}}=\sqrt{\frac{45}{812}}g.$
This is the Coleman-Weinberg condition that guarantees that the new
minimum $\sigma_{1}=\eta$ is in the range of the perturbative calculations
of our approach. In the renormalization process the constant $X\left(e,\, g\right)$
and the finite counterterm $B_{fin}$ get automatically fixed and
disappear from the expression of $U_{ren}$. The dependence on logaritms
of the coupling constants (present in $X$) completely disappeared
from the result. The renormalized effective potential only relies
on $Y$ which is a polynomial in the coupling constants. The result
is
\begin{equation}
U_{ren}=\frac{Y^{2}}{2}\sigma_{1}^{6}\ln^{2}\left[\frac{\sigma_{1}^{2}}{\eta^{2}}\right].
\end{equation}
The new minimum $\sigma_{1}=\eta$ implies in $\sigma_{2}=0=U_{ren}.$
This result means that supersymmetry is preserved but the gauge $U\left(1\right)$
symmetry is broken through a Higgs mechanism that is radiatively induced.\foreignlanguage{english}{
In order to analyse the spectrum of the resulting quantum excitations
we now restore the rotational symmetry by performing the substitution
$\sigma_{1}^{2}\rightarrow\sigma_{1}^{2}+\pi_{1}^{2}$. The above
potential becomes
\begin{equation}
U_{ren}(\sigma_{1},\,\pi_{1})=\frac{1}{2}\left[\frac{e^{6}}{128\pi^{2}}\right]^{2}(\sigma_{1}^{2}+\pi_{1}^{2})^{3}\,\ln^{2}\left[(\sigma_{1}^{2}+\pi_{1}^{2})/\eta^{2}\right].
\end{equation}
A continuous set of new vacua are given by $\sigma_{1}^{2}+\pi_{1}^{2}=\eta^{2}.$
Let us choose the vacuum $\sigma_{1}=\eta$ and $\pi_{1}=0$. The
quantum fields around this new vacuum present a Higgs mechanism \cite{Higgs}.
The mass of the Higgs superfield $\Sigma$ and the Goldstone superfield
$\Pi$ are got from the second derivatives of the effective potential
at the vacuum:
\begin{equation}
m_{\Sigma}^{2}=\left.\frac{\partial^{2}U_{ren}}{\partial\sigma_{1}^{2}}\right|_{(\sigma_{1},\pi_{1})=(\eta,0)}=4\,\left[\frac{e^{6}}{128\pi^{2}}\right]^{2}\eta^{4}
\end{equation}
\begin{equation}
m_{\Pi}^{2}=\left.\frac{\partial^{2}U_{ren}}{\partial\pi_{1}^{2}}\right|_{(\sigma_{1},\pi_{1})=(\eta,0)}=0
\end{equation}
The mass generation for the gauge superfield $A_{\alpha}$ can be
seen in the following way. After renormalization and restoration of
the rotational symmetry, (\ref{eq:3.5}) becomes
\begin{equation}
U_{ren}=-\frac{1}{2}(\sigma_{2}^{2}+\pi_{2}^{2})-Y(\sigma_{1}\sigma_{2}+\pi_{1}\pi_{2})(\sigma_{1}^{2}+\pi_{1}^{2})\ln\left[(\sigma_{1}^{2}+\pi_{1}^{2})/\eta^{2}\right].\label{eq:3.8}
\end{equation}
As shown above, the first term comes from the kinetic terms of $\Sigma$
and $\Pi$ in the action of  (\ref{eq:2.5}). The second term replaces
the classical interaction potential $U_{cl}=g(\sigma_{1}^{2}+\pi_{1}^{2})(\sigma_{1}\sigma_{2}+\pi_{1}\pi_{2})$
that, in turn, comes from the term
\begin{equation}
\delta S_{cl}=-\int d^{5}z\,\frac{g}{4}\left(\Sigma^{2}+\Pi^{2}\right)^{2}\label{eq:3.9}
\end{equation}
 in (\ref{eq:2.5}). In the same way, the second term of $U_{ren}$
in (\ref{eq:3.8}) can be obtained from 
\begin{equation}
\delta S_{eff}=\int d^{5}z\,\frac{Y}{4}\,(\Sigma^{2}+\Pi^{2})^{2}\left\{ \ln\left[(\Sigma^{2}+\Pi^{2})/\eta^{2}\right]+\frac{1}{2}\right\} ,\label{eq:3.10}
\end{equation}
after shifting the fields by their classical expectation values $\sigma$
and $\pi$ and integrating in $d^{2}\theta$. }

\selectlanguage{english}%
The effect of the radiative corrections is to change the classical
potential by the effective one. Forgetting other possible radiative
corrections to the kinetic terms, the effective action is then given
by (\ref{eq:2.5}) with the classical interaction potential (\ref{eq:3.9})
substituted by the effective one (\ref{eq:3.10}). By doing the shift
$\Sigma\longrightarrow\Sigma+\eta$ in this effective action, we see
that a mass term $m_{A}A^{\alpha}A_{\alpha}/2$ with $m_{A}=e^{2}\eta^{2}/2$
is induced for the gauge superfield (besides the mass term $-\frac{1}{2}m_{\Sigma}\Sigma^{2}$
for $\Sigma$). Yet, a bilinear mixing term of the form $\frac{e}{2}\eta\Pi D_{\alpha}A^{\alpha}$
is also induced in the action. These two facts are features of the
Higgs mechanism \cite{Higgs}: the gauge field combines with the ``would-be''
Goldstone scalar superfield, absorbing its degrees of freedom and
becoming massive. In our case the originally non propagating gauge
superfield $A_{\alpha}$ absorbs the degrees of freedom of the super-Goldstone
field $\Pi$, becoming a massive propagating superfield. 

\selectlanguage{american}%

\section{SUMMARY AND CONCLUSIONS}

In this paper the effective potential up to two loops (in the Landau
gauge $\alpha\rightarrow0$ and $\sigma_{2}$ linear approximation)
of the $\mathcal{N}=1$ supersymmetric Chern-Simons model minimally
coupled to matter (SCSM$_{3}$) is calculated by using the tadpole
\cite{weinberg-1973} (for one loop calculations) and the vacuum bubble
\cite{jackiw} (for two loops) methods in the superfield formalism.
In these methods, the scalar superfields have to be shifted by their
$\theta$ dependent vacuum expectation values, breaking explicitly
the supersymmetry in the intermediate stages of the calculation. In
order to derive the superpropagators of the broken susy SCSM$_{3}$
model (the shifted theory) we have employed the projection operator
method developed in \cite{boldo-helayel} and recently enlarged (in
the mixing and gauge sectors) in \cite{gallegos-adilson}. By analyzing
the minimum of the two-loop effective potential, we conclude that
supersymmetry is preserved under radiative corrections, while the
internal $U\left(1\right)$ symmetry is dynamically broken at two-loop
level, generating masses both for the gauge superfield $A_{\alpha}$
and for the matter scalar (Higgs) superfield $\Sigma$. As supersymmetry
is preserved, the masses of the bosonic and fermionic component fields
for each one of the superfields are the same. \foreignlanguage{english}{The
ratio of the induced masses is $m_{\Sigma}^{2}/m_{A}^{2}=(e^{4}/32\pi^{2})^{2}.$ }

\begin{center}
\textbf{ACKNOWLEDGMENTS} 
\par\end{center}

This work was partially supported by the Brazilian agencies Conselho
Nacional de Desenvolvimento Cientifico e Tecnológico (CNPq) and Fundação
de Amparo à Pesquisa do Estado de São Paulo (FAPESP). The authors
would like to thank A. F. Ferrari for the implementation of the SusyMath
package to the case of explicit broken supersymmetric theories in
3D.

\appendix

\section{THE SUPERPROPAGATOR COEFFICIENTS \label{sec:App-Coeff}}

In this Appendix we list the coefficients of the superpropagators
of the shifted Coleman-Weinberg model. These were derived, in the
$\left(\alpha,\,\sigma_{2}\right)$-linear approximation, by using
the projection operator method developed in \cite{boldo-helayel}
and enlarged in \cite{gallegos-adilson}. 

The gauge superpropagator $\left\langle AA\right\rangle $ is given
by

\begin{equation}
\left\langle A_{\alpha}\left(k,\,\theta\right)A_{\beta}\left(-k,\,\theta^{\prime}\right)\right\rangle =i\left\{ \mbox{\ensuremath{\ensuremath{\sum\limits _{i=0}^{5}\left(r_{i}R_{i,\alpha\beta}+s_{i}S_{i,\alpha\beta}\right)}+}\ensuremath{m}}M_{\alpha\beta}+nN_{\alpha\beta}\right\} \delta^{2}\left(\theta-\theta^{\prime}\right),
\end{equation}
with
\begin{eqnarray*}
r_{0} & = & -\frac{\alpha}{2k^{2}}-\frac{\sigma_{1}^{5}\sigma_{2}e^{6}+64k^{4}+16k^{2}\left(4\mu_{1}^{2}-e^{2}\sigma_{1}\sigma_{2}\right)}{256k^{2}\left(k^{2}+\mu_{1}^{2}\right)^{2}}\\
r_{1} & = & \frac{\alpha e^{2}\left(e^{2}\mu_{2}^{2}-4k^{2}g\right)\sigma_{1}^{3}\sigma_{2}}{16k^{4}\left(k^{2}+\mu_{1}^{2}\right)\left(k^{2}+\mu_{2}^{2}\right)}-\frac{e^{2}\sigma_{1}^{2}\left(\sigma_{1}\sigma_{2}e^{2}-2k^{2}-2\mu_{1}^{2}\right)}{32k^{2}\left(k^{2}+\mu_{1}^{2}\right)^{2}}\\
r_{2} & = & -\frac{e^{4}\sigma_{1}^{3}\sigma_{2}}{16\left(k^{2}+\mu_{1}^{2}\right)^{2}}=2s_{3}=s_{4}\\
r_{3} & = & \frac{e^{2}\left(\mu_{1}^{2}-k^{2}\right)\sigma_{1}\sigma_{2}}{16k^{2}\left(k^{2}+\mu_{1}^{2}\right)^{2}}=\frac{1}{2}r_{4}=-\frac{1}{2k^{2}}s_{2}\\
r_{5} & = & \frac{\alpha e^{2}\left(e^{2}\mu_{2}^{2}-4k^{2}g\right)\sigma_{1}^{3}\sigma_{2}}{16k^{4}\left(k^{2}+\mu_{1}^{2}\right)\left(k^{2}+\mu_{2}^{2}\right)}-\frac{e^{4}\sigma_{1}^{3}\sigma_{2}}{32k^{2}\left(k^{2}+\mu_{1}^{2}\right)^{2}}\\
s_{0} & = & \frac{e^{2}\sigma_{1}^{2}\left(\sigma_{1}\sigma_{2}e^{2}-2k^{2}-2\mu_{1}^{2}\right)}{32\left(k^{2}+\mu_{1}^{2}\right)^{2}}\\
s_{1} & = & \frac{\alpha\left(e^{2}g\left(e^{2}\left(g\sigma_{1}^{3}-2\sigma_{2}\right)-8g\sigma_{2}\right)\sigma_{1}^{5}+k^{2}\left(e^{4}+16g^{2}\right)\sigma_{1}^{4}+16k^{4}\right)}{32k^{2}\left(k^{2}+\mu_{1}^{2}\right)\left(k^{2}+\mu_{2}^{2}\right)}\\
 &  & -\frac{\sigma_{1}^{5}\sigma_{2}e^{6}+64k^{4}+16k^{2}\left(4\mu_{1}^{2}-e^{2}\sigma_{1}\sigma_{2}\right)}{256k^{2}\left(k^{2}+\mu_{1}^{2}\right)^{2}}\\
s_{5} & = & -\frac{\alpha e^{2}g\left(e^{2}+4g\right)\sigma_{2}\sigma_{1}^{5}}{16k^{2}\left(k^{2}+\mu_{1}^{2}\right)\left(k^{2}+\mu_{2}^{2}\right)}-\frac{e^{2}\left(\mu_{1}^{2}-k^{2}\right)\sigma_{2}\sigma_{1}}{16k^{2}\left(k^{2}+\mu_{1}^{2}\right)^{2}}\\
m & \sim & \mathcal{O}(\alpha^{2})\qquad\qquad n=0.
\end{eqnarray*}
Here the masses $\mu_{1}$, $\mu_{2}$, $\mu_{3}$ are defined by
the relations $4\mu_{1}\doteq e^{2}\sigma_{1}^{2}$ and $3\mu_{2}\doteq\mu_{3}\doteq3g\,\sigma_{1}^{2}$.

The scalar superpropagator $\left\langle \Pi\Pi\right\rangle $ is
given by
\begin{equation}
\left\langle \Pi\left(k,\theta\right)\Pi\left(-k,\theta^{\prime}\right)\right\rangle =i\,\left(\sum\limits _{i=0}^{5}a_{i}P_{i}\right)\delta^{2}\left(\theta-\theta^{\prime}\right),
\end{equation}
where
\begin{eqnarray*}
a_{0} & = & \frac{\alpha e^{2}\sigma_{1}^{2}\left(k^{4}+g^{3}\sigma_{1}^{5}\left(8\sigma_{2}-g\sigma_{1}^{3}\right)\right)}{2\left(k^{2}+\mu_{2}^{2}\right)^{3}}-\frac{g\sigma_{1}^{2}\left(k^{2}+g\sigma_{1}\left(g\sigma_{1}^{3}-2\sigma_{2}\right)\right)}{\left(k^{2}+\mu_{2}^{2}\right)^{2}}\\
a_{1} & = & \frac{\alpha e^{2}\sigma_{1}\left(k^{2}\left(\sigma_{2}-g\sigma_{1}^{3}\right)+\mu_{2}^{2}\left(5\sigma_{2}-g\sigma_{1}^{3}\right)\right)}{\left(k^{2}+\mu_{2}^{2}\right)^{3}}-\frac{k^{2}+g\sigma_{1}\left(g\sigma_{1}^{3}-2\sigma_{2}\right)}{\left(k^{2}+\mu_{2}^{2}\right)^{2}}\\
a_{2} & = & \frac{2\sigma_{1}\sigma_{2}\left(g\, k^{2}-g^{3}\sigma_{1}^{4}\right)}{\left(k^{2}+\mu_{2}^{2}\right)^{2}}-\frac{e^{2}\alpha\sigma_{1}\sigma_{2}\left(k^{4}-6k^{2}\mu_{2}^{2}+g^{4}\sigma_{1}^{8}\right)}{\left(k^{2}+\mu_{2}^{2}\right)^{3}}\\
a_{3} & = & \frac{2\alpha ge^{2}\sigma_{1}^{3}\sigma_{2}\left(k^{2}-\mu_{2}^{2}\right)}{\left(k^{2}+\mu_{2}^{2}\right)^{3}}-\frac{2g^{2}\sigma_{1}^{3}\sigma_{2}}{\left(k^{2}+\mu_{2}^{2}\right)^{2}}=\frac{1}{2}a_{4}\\
a_{5} & = & \frac{\alpha e^{2}g^{2}\sigma_{1}^{5}\sigma_{2}\left(5k^{2}+\mu_{2}^{2}\right)}{k^{2}\left(k^{2}+\mu_{2}^{2}\right)^{3}}+\frac{2g\sigma_{1}\sigma_{2}}{\left(k^{2}+\mu_{2}^{2}\right)^{2}}.
\end{eqnarray*}

The mixing superpropagator $\left\langle \Pi A\right\rangle $ exhibits
the following structure
\begin{equation}
\left\langle T\,\Pi\left(k,\,\theta\right)A_{\alpha}\left(-k,\,\theta^{\prime}\right)\right\rangle =i\left(\sum_{i=1}^{8}b_{i}\, T_{\alpha}^{i}\right)\delta^{2}\left(\theta-\theta^{\prime}\right),
\end{equation}
where
\begin{eqnarray*}
b_{1} & = & -\frac{e\alpha\sigma_{2}\sigma_{1}^{4}\left(k^{2}g\left(8g-e^{2}\right)+e^{2}\mu_{2}^{2}\left(e^{2}+g\right)\right)}{8\left(k^{2}+\mu_{1}^{2}\right)\left(k^{2}+\mu_{2}^{2}\right)^{2}}-\frac{e\sigma_{2}\left(e^{2}g\sigma_{1}^{4}-4k^{2}\right)}{16\left(k^{2}+\mu_{1}^{2}\right)\left(k^{2}+\mu_{2}^{2}\right)},\\
b_{2} & = & -\frac{e\sigma_{2}\sigma_{1}^{2}\left(e^{2}+4g\right)}{16\left(k^{2}+\mu_{1}^{2}\right)\left(k^{2}+\mu_{2}^{2}\right)}-\frac{e\alpha g\sigma_{2}\sigma_{1}^{2}\left(\sigma_{1}^{4}\left(e^{4}+2e^{2}g+4g^{2}\right)+12k^{2}\right)}{8\left(k^{2}+\mu_{1}^{2}\right)\left(k^{2}+\mu_{2}^{2}\right)^{2}}\\
b_{3} & = & \frac{e\alpha g\sigma_{1}^{2}\sigma_{2}\left(e^{4}g^{2}\sigma_{1}^{8}+k^{2}\sigma_{1}^{4}\left(e^{4}-4e^{2}g+8g^{2}\right)+24k^{4}\right)}{16k^{2}\left(k^{2}+\mu_{1}^{2}\right)\left(k^{2}+\mu_{2}^{2}\right)^{2}}-\frac{e\sigma_{1}^{2}\sigma_{2}\left(e^{2}+4g\right)}{16\left(k^{2}+\mu_{1}^{2}\right)\left(k^{2}+\mu_{2}^{2}\right)}\\
b_{4} & = & \frac{e\alpha g\sigma_{2}\sigma_{1}^{4}\left(e^{2}\mu_{2}^{2}-k^{2}\left(e^{2}+8g\right)\right)}{8k^{2}\left(k^{2}+\mu_{1}^{2}\right)\left(k^{2}+\mu_{2}^{2}\right)^{2}}+\frac{e\sigma_{2}\left(e^{2}g\sigma_{1}^{4}-4k^{2}\right)}{16k^{2}\left(k^{2}+\mu_{1}^{2}\right)\left(k^{2}+\mu_{2}^{2}\right)},\\
b_{5} & = & -\frac{e\alpha\sigma_{1}\left(k^{2}+g\sigma_{1}\left(g\sigma_{1}^{3}-2\sigma_{2}\right)\right)}{2\left(k^{2}+\mu_{2}^{2}\right)^{2}},\\
b_{6} & = & \frac{e\alpha\left(k^{2}\left(g\sigma_{1}^{3}-\sigma_{2}\right)+\mu_{2}^{2}\left(g\sigma_{1}^{3}-3\sigma_{2}\right)\right)}{2k^{2}\left(k^{2}+\mu_{2}^{2}\right)^{2}}\\
b_{7} & = & \frac{e\alpha\sigma_{2}\left(k^{2}-3\mu_{2}^{2}\right)}{2\left(k^{2}+\mu_{2}^{2}\right)^{2}}\\
b_{8} & = & \frac{e\alpha g\sigma_{1}^{2}\sigma_{2}\left(\mu_{2}^{2}-3k^{2}\right)}{2k^{2}\left(k^{2}+\mu_{2}^{2}\right)^{2}}.
\end{eqnarray*}

Finally, the scalar superpropagator $\left\langle \Sigma\Sigma\right\rangle $
is given by
\begin{equation}
\left\langle \Sigma\left(k,\theta\right)\Sigma\left(-k,\theta^{\prime}\right)\right\rangle =i\,\left(\sum\limits _{i=0}^{5}c_{i}P_{i}\right)\delta^{2}\left(\theta-\theta^{\prime}\right),
\end{equation}
with
\begin{eqnarray*}
c_{0} & = & \frac{3g\sigma_{1}^{2}\left(-k^{2}+6g\sigma_{1}\sigma_{2}-\mu_{3}^{2}\right)}{\left(k^{2}+\mu_{3}^{2}\right)^{2}}\\
c_{1} & = & -\frac{k^{2}-6g\sigma_{1}\sigma_{2}+\mu_{3}^{2}}{\left(k^{2}+\mu_{3}^{2}\right)^{2}}\\
c_{2} & = & \frac{6\sigma_{2}\left(k^{2}g\sigma_{1}-9g^{3}\sigma_{1}^{5}\right)}{\left(k^{2}+\mu_{3}^{2}\right)^{2}}\\
c_{3} & = & -\frac{18g^{2}\sigma_{1}^{3}\sigma_{2}}{\left(k^{2}+\mu_{3}^{2}\right)^{2}}=\frac{1}{2}c_{4},\qquad c_{5}=\frac{6g\sigma_{1}\sigma_{2}}{\left(k^{2}+\mu_{3}^{2}\right)^{2}}.
\end{eqnarray*}

\section{TWO-LOOP CALCULATIONS \label{sec:App-2-loops}}

The Feynman diagrams which contribute to the effective potential of
the Coleman-Weinberg at the two-loop order, in the vacuum bubble method,
are depicted in Figure \ref{fig:2-loop}. After performing the integration
over the $\theta$ variables (i.e. the D-algebra) through the SusyMath
package \cite{Ferrari}, we obtain the following results (in the Landau
gauge $\alpha=0$ and in the $\sigma_{2}$ linear approximation):
\begin{eqnarray}
U_{2(a)} & = & \frac{1}{2}\int\frac{d^{3}k}{\left(2\pi\right)^{3}}\frac{d^{3}q}{\left(2\pi\right)^{3}}\Biggl[-\frac{e^{8}\sigma_{1}^{11}\sigma_{2}g^{3}k\cdot q\left[9\left(g^{2}\sigma_{1}^{4}+k\cdot q\right)+5q^{2}\right]}{8k^{2}\left(k^{2}+\mu_{1}^{2}\right)^{2}\left(q^{2}+\mu_{3}^{2}\right)^{2}\left[(k+q)^{2}+\mu_{2}^{2}\right]^{2}}\nonumber \\
 &  & +\frac{e^{2}\sigma_{1}^{3}\sigma_{2}}{8\left(k^{2}+\mu_{1}^{2}\right)^{2}\left(q^{2}+\mu_{3}^{2}\right)^{2}\left[(k+q)^{2}+\mu_{2}^{2}\right]^{2}}\Bigl[e^{2}\left(8g-3e^{2}\right)k\cdot q\, q^{4}-144g^{2}k\cdot q\, k^{4}\nonumber \\
 &  & -\left(e^{4}-12ge^{2}+112g^{2}\right)k^{2}q^{4}-9e^{2}g^{3}\left(2e^{4}+9ge^{2}+8g^{2}\right)\sigma_{1}^{8}k\cdot q-288g^{2}(k\cdot q)^{2}k^{2}\nonumber \\
 &  & +12g\left(e^{2}-8g\right)k^{4}q^{2}-2e^{2}\left(e^{2}-8g\right)(k\cdot q)^{2}q^{2}-\left(e^{4}-32ge^{2}+352g^{2}\right)k\cdot q\, k^{2}q^{2}\nonumber \\
 &  & -18g^{2}\left(e^{4}+16g^{2}\right)\sigma_{1}^{4}k\cdot q\, k^{2}-56e^{4}g^{2}\sigma_{1}^{4}k\cdot q\, q^{2}-18g^{2}\left(e^{4}+16g^{2}\right)\sigma_{1}^{4}k^{2}q^{2}-e^{4}q^{6}\nonumber \\
 &  & -36e^{4}g^{2}\sigma_{1}^{4}(k\cdot q)^{2}-20e^{4}g^{2}\sigma_{1}^{4}q^{4}-\frac{3}{2}e^{4}g^{3}\left(5e^{2}+38g\right)\sigma_{1}^{8}\, q^{2}-\frac{27}{2}e^{4}g^{5}\left(e^{2}+4g\right)\sigma_{1}^{12}\nonumber \\
 &  & -\frac{27}{4}g^{3}\left(e^{6}+4ge^{4}+16g^{2}e^{2}+64g^{3}\right)\sigma_{1}^{8}k^{2}\Bigr]\Biggr]
\end{eqnarray}

\begin{eqnarray}
U_{2(e)} & = & \frac{1}{2}\int\frac{d^{3}k}{\left(2\pi\right)^{3}}\frac{d^{3}q}{\left(2\pi\right)^{3}}\Biggl[\frac{27e^{16}g^{3}\sigma_{1}^{19}\sigma_{2}\left(k^{2}+k\cdot q\right)}{16384k^{2}(k+q)^{2}\left(k^{2}+\mu_{1}^{2}\right)^{2}\left(q^{2}+\mu_{3}^{2}\right)^{2}\left[(k+q)^{2}+\mu_{1}^{2}\right]^{2}}\nonumber \\
 &  & -\frac{9e^{14}g^{2}\sigma_{1}^{15}\sigma_{2}\left[\left(k^{2}+k\cdot q\right)k^{2}+k\cdot q(k+q)^{2}+3\left(1+4g/e^{2}\right)k^{2}(k+q)^{2}\right]}{2048k^{2}(k+q)^{2}\left(k^{2}+\mu_{1}^{2}\right)^{2}\left(q^{2}+\mu_{3}^{2}\right)^{2}\left[(k+q)^{2}+\mu_{1}^{2}\right]^{2}}\nonumber \\
 &  & -\frac{9e^{10}g^{2}\sigma_{1}^{11}\sigma_{2}\left[k^{6}+3\left(1+4g/e^{2}\right)k^{2}(k+q)^{4}+(k+q)^{4}k\cdot q+k^{4}k\cdot q\right]}{256k^{2}(k+q)^{2}\left(k^{2}+\mu_{1}^{2}\right)^{2}\left(q^{2}+\mu_{3}^{2}\right)^{2}\left[(k+q)^{2}+\mu_{1}^{2}\right]^{2}}\nonumber \\
 &  & -\frac{e^{12}\sigma_{2}\sigma_{1}^{11}q^{2}\left[3\left(e^{2}+8g\right)k^{2}(k+q)^{2}+\left(e^{2}+12g\right)\left[k^{4}+k\cdot q\left(k^{2}+(k+q)^{2}\right)\right]\right]}{4096k^{2}(k+q)^{2}\left(k^{2}+\mu_{1}^{2}\right)^{2}\left(q^{2}+\mu_{3}^{2}\right)^{2}\left[(k+q)^{2}+\mu_{1}^{2}\right]^{2}}\nonumber \\
 &  & +\frac{e^{6}\sigma_{2}\sigma_{1}^{3}q^{2}\left[k^{4}+3\left(1+4g/e^{2}\right)k^{2}(k+q)^{2}+k\cdot q\left(k^{2}+(k+q)^{2}\right)\right]}{16\left(k^{2}+\mu_{1}^{2}\right)^{2}\left(q^{2}+\mu_{3}^{2}\right)^{2}\left[(k+q)^{2}+\mu_{1}^{2}\right]^{2}}\nonumber \\
 &  & -\frac{3ge^{8}\sigma_{1}^{7}\sigma_{2}\left[3g\left(e^{2}+6g\right)\sigma_{1}^{4}\left(2k^{2}+k\cdot q\right)+4q^{2}\left(k^{2}+k\cdot q\right)\right]}{128\left(k^{2}+\mu_{1}^{2}\right)^{2}\left(q^{2}+\mu_{3}^{2}\right)^{2}\left[(k+q)^{2}+\mu_{1}^{2}\right]^{2}}\Biggr]
\end{eqnarray}

\begin{eqnarray}
U_{2(g)} & = & \frac{1}{2}\int\frac{d^{3}k}{\left(2\pi\right)^{3}}\frac{d^{3}q}{\left(2\pi\right)^{3}}\Biggl[\frac{17496g^{7}\sigma_{1}^{11}\sigma_{2}\left[k^{2}+q^{2}+(k+q)^{2}\right]}{\left(k^{2}+\mu_{3}^{2}\right)^{2}\left(q^{2}+\mu_{3}^{2}\right)^{2}\left[(k+q)^{2}+\mu_{3}^{2}\right]^{2}}\nonumber \\
 &  & +\frac{314928g^{9}\sigma_{1}^{15}\sigma_{2}-216g^{3}\sigma_{1}^{3}\sigma_{2}k^{2}q^{2}(k+q)^{2}}{\left(k^{2}+\mu_{3}^{2}\right)^{2}\left(q^{2}+\mu_{3}^{2}\right)^{2}\left[(k+q)^{2}+\mu_{3}^{2}\right]^{2}}\Biggr]
\end{eqnarray}

\begin{eqnarray}
U_{2(h)} & = & \frac{1}{2}\int\frac{d^{3}k}{\left(2\pi\right)^{3}}\frac{d^{3}q}{\left(2\pi\right)^{3}}\Biggl[\frac{40g^{7}\sigma_{1}^{11}\sigma_{2}\left(9k^{2}+q^{2}+9(k+q)^{2}\right)}{\left(k^{2}+\mu_{2}^{2}\right)^{2}\left(q^{2}+\mu_{3}^{2}\right)^{2}\left[(k+q)^{2}+\mu_{2}^{2}\right]^{2}}\nonumber \\
 &  & +\frac{720g^{9}\sigma_{2}\sigma_{1}^{15}-40g^{3}\sigma_{2}\sigma_{1}^{3}k^{2}q^{2}(k+q)^{2}}{\left(k^{2}+\mu_{2}^{2}\right)^{2}\left(q^{2}+\mu_{3}^{2}\right)^{2}\left[(k+q)^{2}+\mu_{2}^{2}\right]^{2}}\Biggr]
\end{eqnarray}

\begin{eqnarray}
U_{2(i)} & =\frac{1}{32} & \int\frac{d^{3}k}{\left(2\pi\right)^{3}}\frac{d^{3}q}{\left(2\pi\right)^{3}}\frac{18e^{6}g^{2}\sigma_{1}^{7}\sigma_{2}+e^{2}\sigma_{1}^{3}\sigma_{2}\left(e^{4}q^{2}+144g^{2}k^{2}\right)}{\left(k^{2}+\mu_{1}^{2}\right)^{2}\left(q^{2}+\mu_{3}^{2}\right)^{2}}
\end{eqnarray}

\begin{eqnarray}
U_{2(j)} & = & \frac{1}{32}\int\frac{d^{3}k}{\left(2\pi\right)^{3}}\frac{d^{3}q}{\left(2\pi\right)^{3}}\frac{2e^{6}g^{2}\sigma_{1}^{7}\sigma_{2}+e^{2}\sigma_{1}^{3}\sigma_{2}\left(e^{4}q^{2}+16g^{2}k^{2}\right)}{\left(k^{2}+\mu_{1}^{2}\right)^{2}\left(q^{2}+\mu_{2}^{2}\right)^{2}}
\end{eqnarray}

\begin{eqnarray}
U_{2(k)} & = & -\int\frac{d^{3}k}{\left(2\pi\right)^{3}}\frac{d^{3}q}{\left(2\pi\right)^{3}}\frac{486g^{5}\sigma_{1}^{7}\sigma_{2}+27g^{3}\sigma_{1}^{3}\sigma_{2}\left(k^{2}+q^{2}\right)}{\left(k^{2}+\mu_{3}^{2}\right)^{2}\left(q^{2}+\mu_{3}^{2}\right)^{2}}
\end{eqnarray}

\begin{eqnarray}
U_{2(l)} & = & -\int\frac{d^{3}k}{\left(2\pi\right)^{3}}\frac{d^{3}q}{\left(2\pi\right)^{3}}\frac{6g^{5}\sigma_{1}^{7}\sigma_{2}+3g^{3}\sigma_{1}^{3}\sigma_{2}\left(k^{2}+q^{2}\right)}{\left(k^{2}+\mu_{2}^{2}\right)^{2}\left(q^{2}+\mu_{2}^{2}\right)^{2}}
\end{eqnarray}

\begin{eqnarray}
U_{2(m)} & = & -\int\frac{d^{3}k}{\left(2\pi\right)^{3}}\frac{d^{3}q}{\left(2\pi\right)^{3}}\frac{36g^{5}\sigma_{1}^{7}\sigma_{2}+2g^{3}\sigma_{1}^{3}\sigma_{2}\left(9k^{2}+q^{2}\right)}{\left(k^{2}+\mu_{2}^{2}\right)^{2}\left(q^{2}+\mu_{3}^{2}\right)^{2}}
\end{eqnarray}
The other vacuum bubbles which involve the mixing superpropagator
$\left\langle \Pi A\right\rangle $ are null in the Landau gauge ($\alpha=0$).
That is, $U_{2(b)}\sim O\left(\alpha^{2},\,\sigma_{2}\right)$, $U_{2(c)}\sim O\left(\alpha^{2},\,\sigma_{2}\right)$,
$U_{2(d)}\sim O\left(\alpha^{2},\,\sigma_{2}\right)$, $U_{2(f)}\sim O\left(\alpha^{2},\,\sigma_{2}\right)$,
and $U_{2(n)}\sim O\left(\alpha,\,\sigma_{2}^{2}\right)$.

\end{document}